\documentclass[journal]{IEEEtran}

\usepackage{verbatim}
\usepackage{amsfonts}
\usepackage{amssymb}
\usepackage{stfloats}
\usepackage{cite}
\usepackage{graphicx}
\usepackage{psfrag}
\usepackage{amsmath}
\usepackage{array}
\usepackage{epstopdf}
\usepackage{authblk}
\usepackage{graphicx} 
\usepackage{amsthm} 
\usepackage{lipsum}
\usepackage{verbatim} 
\usepackage{authblk}
\usepackage{mathtools}
\usepackage{cuted}
\usepackage{booktabs}

\usepackage{amsmath}
\usepackage{mathrsfs}

\usepackage{algorithmic}

\usepackage{array}

\ifCLASSOPTIONcompsoc
\usepackage[caption=false,font=normalsize,labelfont=sf,textfont=sf]{subfig}
\else
\usepackage[caption=false,font=footnotesize]{subfig}
\fi

\usepackage{url}
\usepackage{graphicx,amsmath,amssymb,amsfonts}
\usepackage{algorithmic,algorithm}

\usepackage{hyperref}
\hypersetup{colorlinks=true, citecolor=blue}

\usepackage{setspace}

\makeatletter

\newcommand{\Rmnum}[1]{\expandafter\@slowromancap\romannumeral #1@}
\makeatother

\usepackage{framed} 

\usepackage{cancel}

\usepackage{booktabs}
\usepackage{tabularx,makecell,multirow}

\usepackage[table]{xcolor}
\usepackage{pifont} 

\usepackage{diagbox}

\definecolor{shadecolor}{rgb}{0.92,0.92,0.92}

\begin{document}
	\bstctlcite{ref:BSTcontrol}

\title{\fontsize{23 pt}{\baselineskip}\selectfont Two Birds With One Stone: Enhancing Communication and Sensing via Multi-Functional RIS}

\author{Wanli Ni, 
	Wen Wang, Ailing Zheng, Peng Wang,
	Changsheng You,
	Yonina C. Eldar, \IEEEmembership{Fellow,~IEEE}, \\
	Dusit Niyato, \IEEEmembership{Fellow,~IEEE},
	and Robert Schober, \IEEEmembership{Fellow,~IEEE}
\vspace{-6 mm}
\thanks{Wanli Ni is with the Department of Electronic Engineering, Tsinghua University, Beijing 100084, China (e-mail: niwanli@tsinghua.edu.cn).}
\thanks{Wen Wang is with the Pervasive Communications Center, Purple Mountain Laboratories, Nanjing 211111, China (email: wangwen@pmlabs.com.cn).}
\thanks{Ailing Zheng is with the State Key Laboratory of Networking and Switching Technology, Beijing University of Posts and Telecommunications, Beijing 100876, China (e-mail: ailing.zheng@bupt.edu.cn).}
\thanks{Peng Wang is with the School of Electrical and Electronic Engineering, North China Electric Power University, Beijing 102206, China (e-mail: wangpeng9712@ncepu.edu.cn).}
\thanks{Changsheng You is with the Department of Electronic and Electrical Engineering, Southern University of Science and Technology, Shenzhen 518055, China (e-mail: youcs@sustech.edu.cn).}
\thanks{Yonina C. Eldar is with the Faculty of Mathematics and Computer Science, Weizmann Institute of Science, Israel (e-mail: yonina.eldar@weizmann.ac.il).}
\thanks{Dusit Niyato is with the College of Computing and Data Science, Nanyang Technological University, Singapore (email: dniyato@ntu.edu.sg).}
\thanks{Robert Schober is with the Institute for Digital Communications, Friedrich-Alexander University of Erlangen-Nuremberg (FAU), 91054 Erlangen, Germany (email: robert.schober@fau.de).}
}

\maketitle
	
	\begin{abstract}
	In this article, we propose new network architectures that integrate multi-functional reconfigurable intelligent surfaces (MF-RISs) into 6G networks to enhance both communication and sensing capabilities.
	Firstly, we elaborate how to leverage MF-RISs for improving communication performance in different communication modes including unicast, mulitcast, and broadcast and for different multi-access schemes.
	Next, we emphasize synergistic benefits of integrating MF-RISs with wireless sensing, enabling more accurate and efficient target detection in 6G networks.
	Furthermore, we present two schemes that utilize MF-RISs to enhance the performance of integrated sensing and communication (ISAC).
	We also study multi-objective optimization to achieve the optimal trade-off between communication and sensing performance.
	Finally, we present numerical results to show the performance improvements offered by MF-RISs compared to conventional RISs in ISAC.
	We also outline key research directions for MF-RIS under the ambition of 6G.
	\end{abstract}
	
	\vspace{-2 mm}
	\section{Introduction}
	In the upcoming 6G era, cutting-edge applications, such as metaverse, high-resolution video, autonomous vehicles, and intelligent robots, necessitate a significant enhancement in both transmission rate and sensing accuracy in the air interface to deliver unprecedented user experiences and performance~\cite{Jiang2024Terahertz}.
	To fulfill these requirements, several improvements are required in next-generation wireless networks, including advanced network architecture, enhanced spectrum utilization, and expanded sensing range.
	
	Traditional sensing and communication systems are designed separately which inevitably causes low network resource utilization.
	To address this, recently, integrated sensing and communication (ISAC) technology has been proposed, which significantly boosts the spectral, energy, and hardware efficiency by sharing sensing and communication resources \cite{Zheng2019Coexistence, Liu2020Joint, Liu2022ISAC}.
	However, ISAC systems still face challenges in practical applications.
	One key issue is the limited sensing and communication range, especially in high-frequency bands, due to blockages and high path loss~\cite{Chepuri2023ISAC}.
	This restricts signal propagation and coverage, impacting system performance.
	Another challenge lies in the efficient management of multiple beams when communication and sensing coexist \cite{Liu2020Joint}.
	Thus, how to design a joint waveform that satisfies both the communication and sensing requirements is a crucial issue to be addressed.
	
\begin{figure*} [t]
		\centering
		\includegraphics[width=6.6 in]{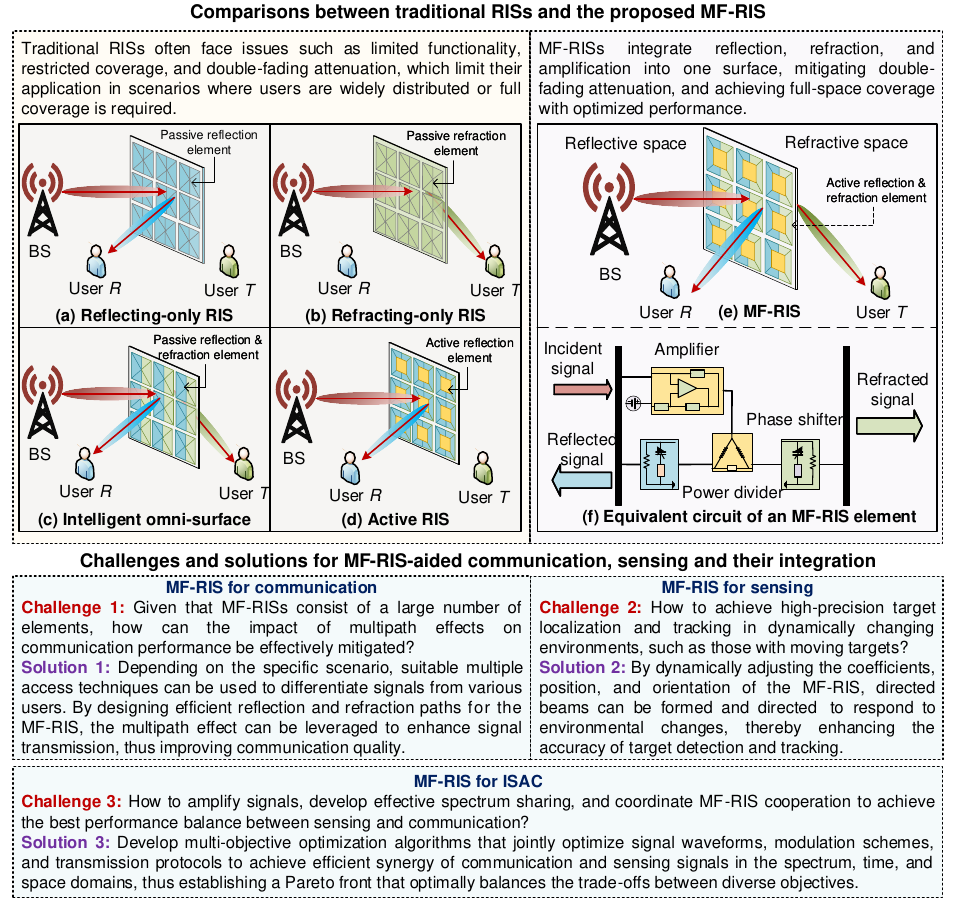}
		\caption{Comparison between MF-RISs and conventional RISs. The hardware circuit of each MF-RIS element comprises an amplifier, a power divider, and two phase shifters. The amplifier and power divider manage the amplitude of the reflected and refracted signals. The phase shifter, which includes multiple PIN diodes, is responsible for adjust the phase shift of the output signal. Besides, we discuss major challenges and solutions for MF-RIS-aided communication, and sensing as well as their integration.}
		\label{Fig1}
	\end{figure*}

	Previous studies have explored the integration of reconfigurable intelligent surfaces (RISs) into ISAC systems to enhance sensing and communication performance simultaneously \cite{Zhang2022Sensing, Liu2023SNR, Shao2024Intelligent}.
	However, conventional RISs face several challenges, such as half-space coverage and the double-fading effect, which hinder their practical implementation.
	To overcome these limitations, we introduce multi-functional RISs (MF-RISs) that support signal reflection, refraction, and amplification simultaneously \cite{Zheng2024TVT, Wang2023IoT, Ni2024RIS}.
	This multifaceted capability helps wireless networks to achieve full-space communication coverage and enhanced sensing capability.
	In Fig. \ref{Fig1}, we compare MF-RISs with conventional RISs.
	Specifically, the signal amplification is particularly critical as it effectively compensate for the double-fading effect, ensuring that signals maintain high strength over long distances.
	Additionally, the ability to reflect and refract signals allows for establishing reliable virtual line-of-sight (LoS) links for both the communication users and sensing targets in the whole space.
	Overall, MF-RISs open new possibilities in ISAC systems, offering robust performance improvements in various ISAC scenarios where traditional RISs suffer degraded performance.
	
	In this article, we investigate the utilization of MF-RISs in communication, sensing, and integrated networks.
	The major contributions are summarized as follows.
	\begin{itemize}
		\item
		\textbf{MF-RIS for communication:}
		MF-RISs optimize wireless communication performance by adjusting their phase shifts and amplitude coefficients.
		In Section~\ref{Section2}, we propose to use MF-RISs to strengthen the received signal power and mitigate interference to boost unicast, multicast, and broadcast communications. We then integrate MF-RISs with multiple access technologies to further improve spectral efficiency.
		
		\item
		\textbf{MF-RIS for sensing:}
		MF-RISs can be used not only for communication but also for wireless sensing, such as target detection and localization.
		To achieve high-precision sensing, we discuss the interplay between MF-RISs and wireless sensing in Section~\ref{Section3}, where we provide MF-RIS-aided solutions for enhanced environmental sensing and target detection.
		
		\item
		\textbf{MF-RIS for ISAC:}
		Waveform design is a key issue in ISAC systems.
		MF-RISs help to achieve efficient data transmission and target sensing with limited radio resources.
		In Section~\ref{Section4}, we present MF-RIS-aided radar-communication coexistence (RCC) and dual-functional radar-communication (DFRC) systems, and emphasize the importance of multi-objective optimization in MF-RIS-aided ISAC systems.
	\end{itemize}

	\begin{figure*} [t]
		\centering
		\includegraphics[width=6.8 in]{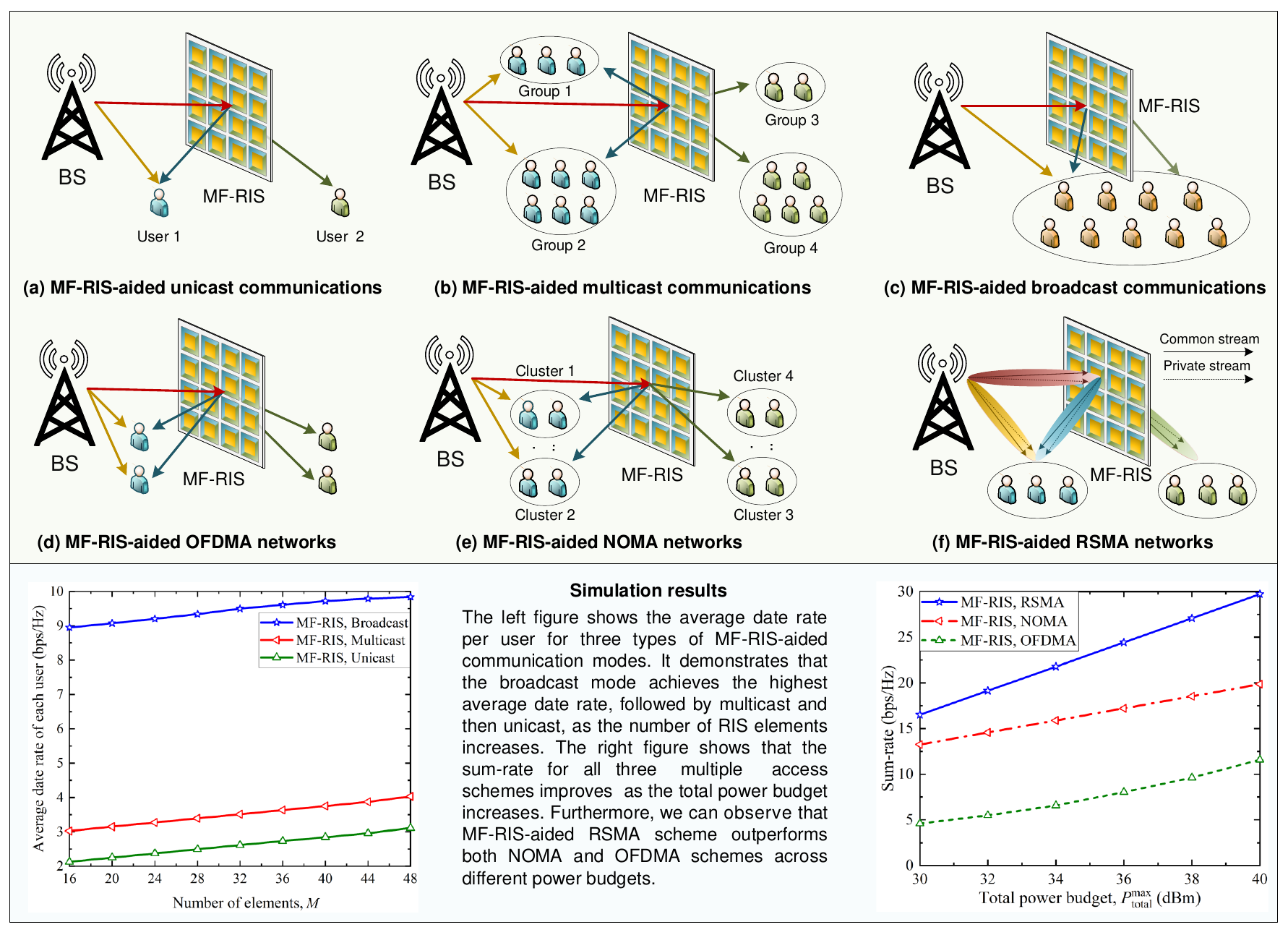}
		\caption{MF-RIS-aided wireless communications and simulation results.}
		\label{Fig2}
	\end{figure*}

	\section{MF-RISs for Communication} \label{Section2}
	
	\subsection{MF-RIS-Aided Communication}
	Data transmission can be performed as unicast, multicast, and broadcast. Specifically, for unicast communication, a base station (BS) sends independent information to each user.
	For multicast communication, the BS sends the same information to multiple users simultaneously \cite{Sadeghi2018Joint}.
	For broadcast communication, the BS sends identical information to all users.
	To further enhance their performance, as shown in Fig. \ref{Fig2}, MF-RISs can be used to assist data distribution.
	
	\subsubsection{MF-RIS-aided unicast communication}
	In a unicast communication system, the transmission of information is carried out in a point-to-point manner, ensuring that the information goes straight from the source to the destination without affecting other nodes that do not need the information.
	Since the information is sent only to a specific recipient, unicast communication empowers the system with a high degree of personalized service, where the BS can respond to specific requests of each user.
	Thanks to its signal amplification capability, MF-RISs can enhance signal strength and mitigate the double-fading attenuation.
	Consequently, applying MF-RISs in unicast communication can significantly improve the individual user experience and boost data transmission efficiency, especially for scenarios where directional data transmission is required \cite{Zheng2023CL}.
	Specifically, according to the real-time channel status of different users, MF-RISs can dynamically adjust their amplitude and phase coefficients to enhance the stability and reliability of unicast links.

	\subsubsection{MF-RIS-aided multicast communication}
	In a multicast communication system, a transmitter (source node) sends information to multiple receivers (destination nodes) at the same time \cite{Sadeghi2018Joint}.
	Compared with unicast communication, multicast communication provides a more efficient data distribution mechanism and significantly reduces network traffic, especially in scenarios where multiple receivers need the same information, such as multiplayer videoconferencing, game synchronization, and software distribution.
	However, the multicast transmission rate is often constrained by the user with the worst channel condition.
	By amplifying incident signals, MF-RISs help enhance the signal reception quality for edge users and reduce power consumption at the BS through their multi-dimensional channel conditioning capabilities.
	Additionally, MF-RISs provide flexible configuration policies, adapting to changes in user distribution within multicast groups.
	For instance, in managing massive devices, MF-RISs' signal manipulation ability allows for the rapid establishment of reliable multicast links.
	This enables the BS to efficiently push control commands or software updates to multiple devices simultaneously, significantly reducing network congestion.
	
	\subsubsection{MF-RIS-aided broadcast communication}
	Unlike unicast and multicast, in a broadcast communication system, data is sent from the source node to all destination nodes in the network simultaneously, regardless of the node location, as long as they are within the network coverage.
	Especially in non-line-of-sight (NLoS) environments, by improving the worst channels in a targeted region, MF-RISs help to achieve efficient transmission of broadcast data and improve the coverage of broadcast services.
	In addition, MF-RISs support multi-direction and multi-beam broadcasting, and dynamically adjust the broadcast strategy according to the importance and urgency of the broadcast content or the receiver's preference, enabling more flexible and customized broadcast services.
	
	In summary, when combining MF-RISs with different communication modes, the fundamental differences lie in how MF-RISs optimize their coefficients to serve users. 
	For unicasting, MF-RISs primarily optimize the reflection/refraction link for each user, tailoring the amplitude and phase shift of each user's signal. 
	For multicasting, it requires a joint optimization of user clustering and MF-RIS configuration to balance among multiple clusters and ensure good signal quality for users within each cluster.
	For broadcasting, the goal of MF-RIS is to cover as large an area as possible so that all users can be served effectively, even at the cell edge.

	\subsection{MF-RIS-Aided Multiple Access}
	In wireless communication systems, multiple access (MA) technologies are continually emerging to meet the increasingly more stringent user requirements, such as increased data rates, lower latency, and higher device density.
	These advancements have evolved from orthogonal frequency division MA (OFDMA) to non-orthogonal MA (NOMA) and further to rate-splitting MA (RSMA).
	As illustrated in Fig. \ref{Fig2}, MF-RISs can be integrated with different MA technologies. 
	
	\subsubsection{MF-RIS-aided OFDMA networks}
	All available spectrum resources are divided into multiple orthogonal subcarriers in OFDMA networks.
	Each subcarrier is independently assigned to different users, which not only supports the simultaneous communication of multiple users, but also greatly improves the spectral efficiency through a fine-grained time-frequency resource allocation strategy.
	Since MF-RISs have more variables to optimize, they provide additional degrees of freedom (DoFs) for signal manipulation compared to conventional RISs.
	Consequently, the performance of OFDMA networks can be significantly enhanced when combined with MF-RISs, even if multiple subcarriers share the same MF-RIS configuration \cite{Wang2023IoT}.

	\subsubsection{MF-RIS-aided NOMA networks}
	NOMA has emerged as a pivotal technology to address the increasing demand for massive connectivity and enhanced spectral efficiency in modern wireless networks.
	This paradigm shift from traditional orthogonal access schemes like OFDMA, allows for a more efficient use of resources by enabling multiple users to simultaneously occupy the same time-frequency resource block.
	NOMA achieves this by employing advanced techniques such as superposition coding at the transmitter, where signals are carefully crafted with different power levels to cater to the varying channel conditions of each user, and successive interference cancellation (SIC) at the receiver, where users with stronger channel conditions decode and subtract the interference from weaker signals before decoding their own \cite{Zheng2023CL}.
	By smartly adjusting the cascaded channels, MF-RISs can create more distinguishable channels that facilitate more efficient SIC, thus enhancing NOMA performance \cite{Zheng2023CL}.
	Simulation results in \cite{Zheng2023CL} revealed that MF-RISs achieve approximately a 50\% improvement in throughput compared to passive RISs.
	
	\subsubsection{MF-RIS-aided RSMA networks}
	RSMA seeks to strike a balance between the efficiency gains of NOMA and the interference management of space-division MA (SDMA). 
	Although NOMA improves spectral efficiency by sharing resources among users, it poses challenges in signal processing complexity and inefficient utilization of multiple antennas, particularly for weak users who must decode all interference signals.
	RSMA tackles this issue by splitting user messages into common and private parts. The common part is decoded by all users, while the private part is for the intended user.
	This allows partial interference cancellation, where some interference is decoded and removed, while the rest is treated as noise.
	By dynamically adjusting the power and content of common and private streams, RSMA can adapt to varying channel conditions, effectively transitioning from SDMA-like behavior when interference is weak to NOMA-like behavior when interference is strong.
	However, the data rate of the common stream is limited by the weakest user's channel.
	Since MF-RISs support signal adjustment with more DoFs compared to conventional RISs, integrating MF-RISs with RSMA can overcome this issue by strategically improving the channel quality for weak users.
	
	In summary, when integrating MF-RISs with different MA schemes, the role of MF-RISs depends on the specific needs. For instance, in OFDMA, the role of MF-RISs is to provide selective phase adjustment for different subcarriers in the frequency domain.
	In NOMA, MF-RISs are primarily used to manage signal superposition and decoding order in power domain.
	In RSMA, MF-RISs are required to simultaneously optimize the transmission of public and private information by balancing inter-user interference.
	These distinctions highlight the adaptive role of MF-RISs in enhancing the performance of different MA schemes.

	\begin{figure*}[t]
		\centering
		\includegraphics[width=6.5 in]{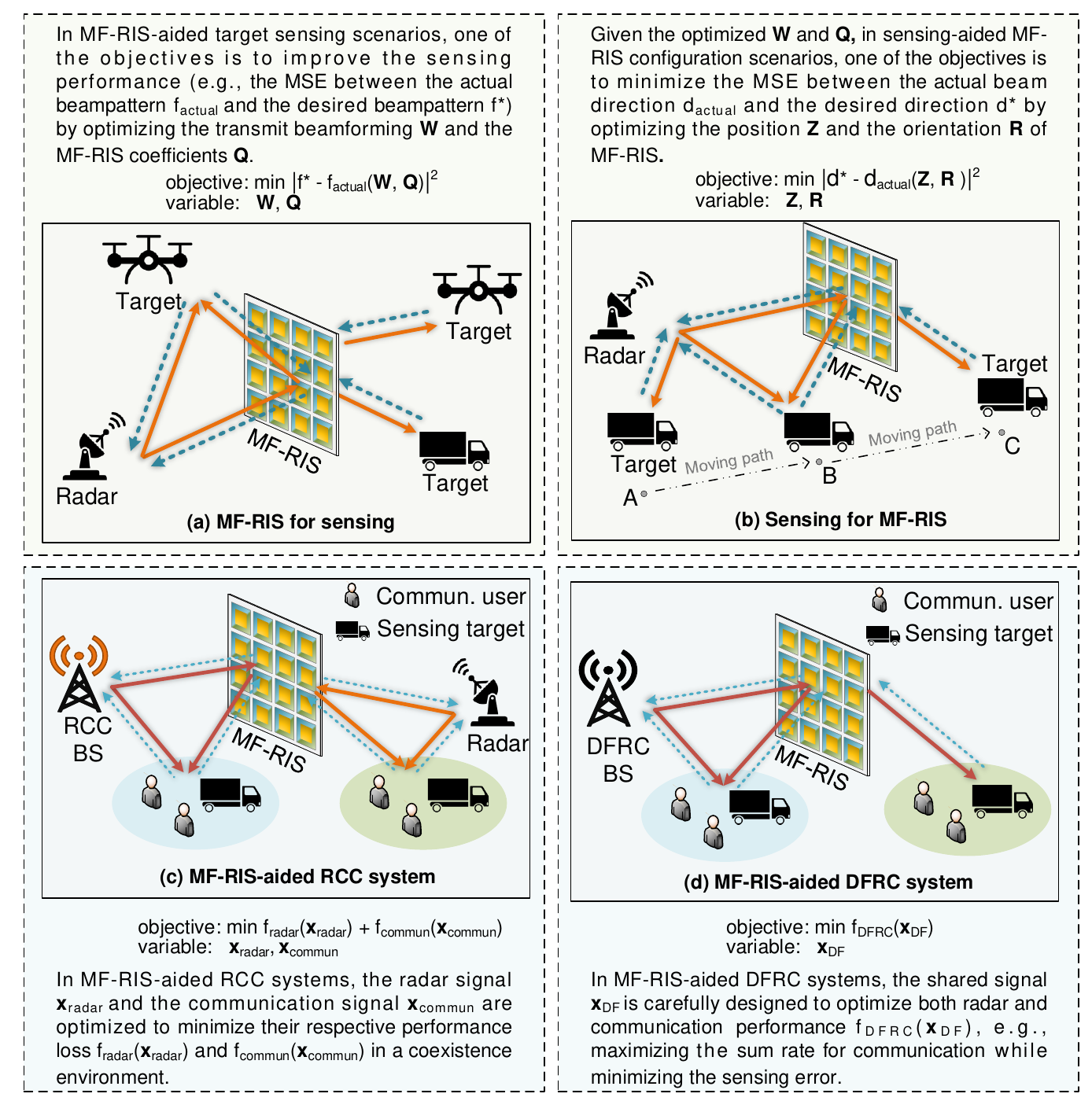}
		\caption{The interplay between MF-RISs and sensing is illustrated in (a) and (b). MF-RIS-aided ISAC systems are illustrated in (c) and (d).}
		\label{Fig3}
	\end{figure*}
    
     \section{Interplay Between MF-RISs and Sensing} \label{Section3}
     In this section, we discuss the interplay between MF-RISs and wireless sensing, exploring how this convergence empowers 6G networks with enhanced sensing capabilities.
        
     \subsection{MF-RIS-Assisted Wireless Sensing}
     	In wireless sensing systems, radar uses radio frequency (RF) signals to detect the environment and targets \cite{Shao2024Intelligent}.
     	However, the low spatial and angular resolution of existing sensing systems often makes it difficult to distinguish signals accurately, especially under NLoS conditions \cite{Elbir2023ISAC}. 
     	In addition, multipath fading and Doppler shift can seriously affect the sensing performance, especially for moving targets.
     	Compared to traditional sensing schemes, MF-RISs offer an innovative solution to address these challenges. 
     	As shown in Fig. \ref{Fig3}(a), the directional beamforming in reflection and refraction spaces of MF-RISs effectively mitigates the interference caused by multipath effects, enabling accurate sensing of targets.
     	Moreover, by reasonably deploying MF-RISs, the echo path (i.e., radar \textrightarrow MF-RIS \textrightarrow target \textrightarrow MF-RIS \textrightarrow radar) can be constructed to create a virtual LoS sensing link.
     	Most importantly, the amplification function of MF-RISs can enhance echo signals, thereby improving both sensing range and accuracy.
     	In addition to optimizing the transmit beamforming at the radar, the coefficients of MF-RIS can be dynamically adjusted to quickly adapt to environmental changes.
     	This real-time control over the amplitude and phase of electromagnetic waves also compensates for Doppler shifts, ensuring stable performance even for fast-moving targets.

     \subsection{Sensing-Assisted MF-RIS Configuration}
     	To clarify the role of wireless sensing in the context of MF-RIS-aided communications, we consider the scenario shown in Fig. \ref{Fig3}(b), where the target progressively approaches the MF-RIS from the far side and then gradually moves away.
     	Initially, when the target is at point A, due to the limited coverage area of MF-RISs, the BS can only sense the target by itself without the help of the MF-RIS.
     	Then, when the target moves from point B to point C, the target trajectory and velocity information sensed at point B can be used to optimize the MF-RIS configuration (e.g., position, rotation, and operating mode \cite{Ni2024RIS}) at point C, thereby proactively preparing the radio environment for the upcoming users and targets.
		Compared to the communication system without sensing assistance, sensing-assisted MF-RIS configuration enables the direction of electromagnetic waves to be more accurately adjusted, and thus the limited energy can be precisely focused on users and targets.
		Finally, when the target moves beyond the MF-RIS’s range, the sensing information gathered by the radar with the assistance of the MF-RIS can further support the subsequent independent sensing of the radar.
	    Gradually, the collaborative sensing between the radar and the MF-RIS becomes more accurate and consumes less resources.
	    This creates a closed-loop feedback mechanism that provides substantial support for sensing in complex scenarios.
	    
	    In summary, the integration of wireless sensing with MF-RIS facilitates high-resolution environmental monitoring, which is beneficial for real-time target sensing and user localization.
	    Additionally, based on sensed data, the network can dynamically adjust the MF-RIS configuration, leading to more efficient use of the radio spectrum.
	    As a result, the bidirectional feedback between sensing and MF-RIS configuration enhances the robustness and reliability under moving targets.

	\begin{figure*}[t]
		\centering
		\includegraphics[width=6.8 in]{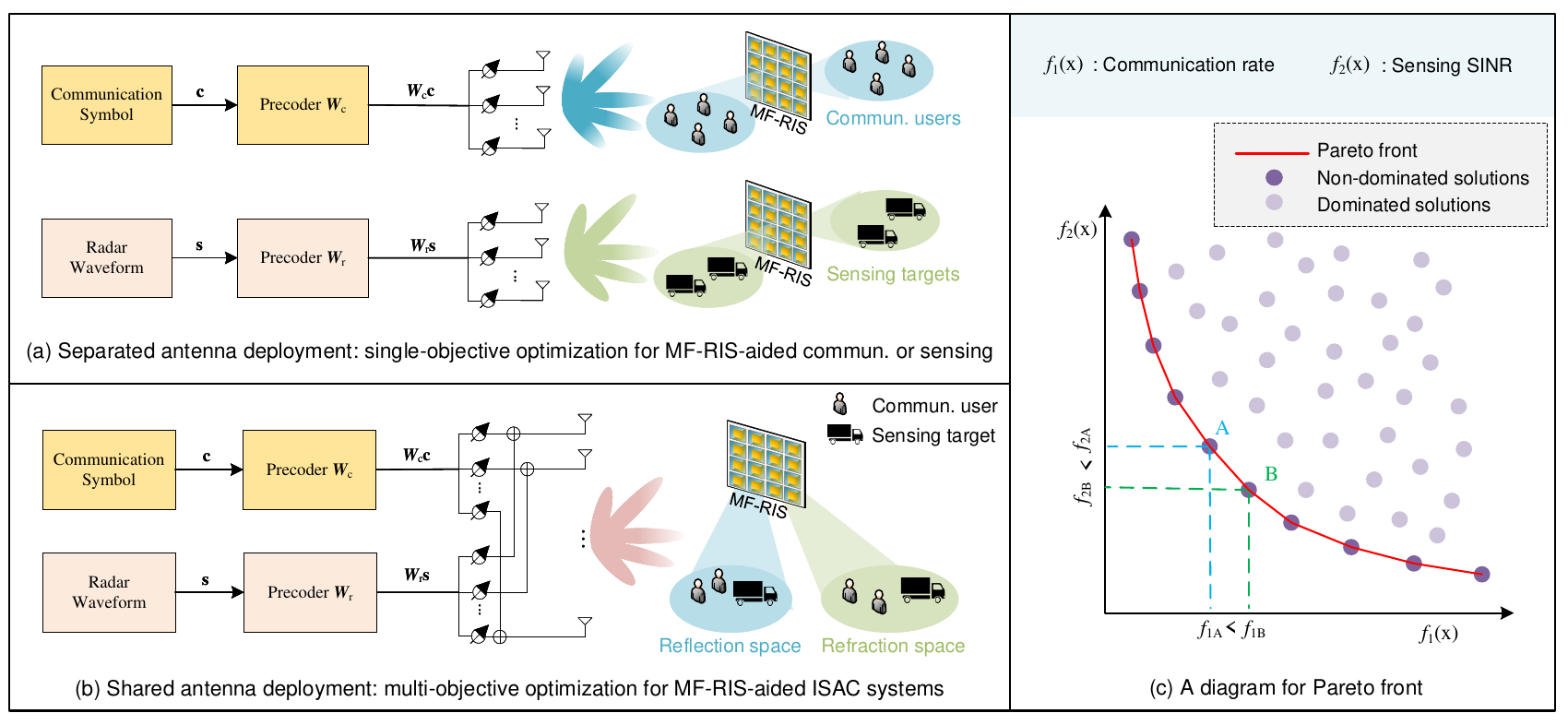}
		\caption{Single-objective optimization versus multi-objective optimization in MF-RIS-aided ISAC systems.}
		\label{Fig4}
	\end{figure*}
	
     \section{MF-RIS-Aided ISAC Systems} \label{Section4}
	 In this section, we first discuss the potential of MF-RISs to enhance ISAC system performance, and then introduce the multi-objective approach for optimizing ISAC systems.
     
     \subsection{Enhancing ISAC Systems with MF-RIS}
     Both RCC and DFRC systems belong to the broader ISAC framework, yet they exhibit distinct operational characteristics.
     RCC employs separate BS and radar hardware for independent sensing and communication \cite{Zheng2019Coexistence}, as shown in Figs. \ref{Fig3}(c).
     Conversely, DFRC integrates both radar and communication functionalities seamlessly within the same BS infrastructure \cite{Liu2022ISAC}, as shown in Figs. \ref{Fig3}(d).
     In addressing the common challenges faced by ISAC systems, several issues have to be addressed.
     Firstly, the shared spectrum resource between sensing and communication signals inevitably leads to interference. 
     Secondly, both sensing and communication signals are susceptible to attenuation and distortion caused by factors such as path loss and multipath effects, which can compromise the quality of both functions.
     Moreover, it is difficult to design dual-purpose waveforms that maximize the communication throughput while minimizing the sensing error.
     Within this context, MF-RISs offer a compelling solution to address these issues. 
     For example, MF-RISs are able to dynamically adjust the waveforms of communication and radar systems. This capability ensures that excessive interference in ISAC systems is effectively mitigated, fostering seamless coexistence \cite{Zhang2022Sensing}.
     Especially for sensing signals, MF-RISs excel in enhancing the reception quality of echo signals. By suppressing interference from other directions, they ensure that the sensing process is unhindered and accurate.
     Again, MF-RISs can leverage its ability to manipulate amplitude and phase shifts to combat the detrimental effects of path loss and multipath on both sensing and communication signals.

     \subsection{Multi-Objective Optimization for ISAC Systems}
     Most previous studies have focused on single-objective optimization that puts one performance (e.g., communication rate) as the objective and another performance metric as a constraint (e.g., sensing accuracy). In the context of ISAC systems, multi-objective optimization is necessary and outperforms single-objective optimization.
     Since ISAC systems integrate communication and sensing functions at the same time, their performance evaluation is not limited to communication quality or sensing accuracy only, but needs to simultaneously consider multiple interrelated and possibly conflicting optimization objectives, as shown in Fig.~\ref{Fig4}.
     However, due to the conflicts among different objectives, multi-objective optimization generally has higher complexity than single-objective optimization which follows a direct approach to deal with only one objective.
     Therefore, balancing multiple objectives simultaneously, such as communication throughput, sensing accuracy, and resource utilization, is crucial for the effective operation of an ISAC system \cite{Wang2024Multi}.
     The aforementioned objectives form a multi-dimensional index system for the performance evaluation of ISAC. 
     For example, in the scenario of autonomous vehicles, the ISAC BS not only provides high-speed and stable communication to facilitate information exchange between vehicles, but also needs to gather crucial data about the surrounding environment through high-precision sensing to ensure driving safety.
     However, these two goals can often conflict with each other in terms of resource allocation.
     Navigating this intricate trade-off requires sophisticated multi-objective optimization techniques (e.g., Pareto optimization, weighted sum method, machine learning, and game theory) to achieve an optimal balance and alleviate resource competition among communication users and sensing targets.
    
    In summary, although MF-RISs can be used to simultaneously improve the reliability of communication links and the accuracy of sensing operations, it also introduces new challenges.
    These include establishing performance limits, effectively amplifying signals, and coordinating the collaboration among MF-RIS elements.
	Specifically, determining the theoretical limits of MF-RIS-aided ISAC systems is essential for setting performance expectations.
	Additionally, designing the dual-purpose waveform helps balance the conflicts between communication and sensing.
	Finally, coordinating MF-RIS elements via mode selection and element allocation enables the ISAC system to dynamically respond to different environments and application needs.

	\begin{figure*} [t!]
		\subfloat[Sum SINR of sensing targets versus the QoS of communication users]{\label{Fig5_2}
			\begin{minipage}[t]{0.45 \textwidth}
				\centering
				\includegraphics[width= 3.2 in]{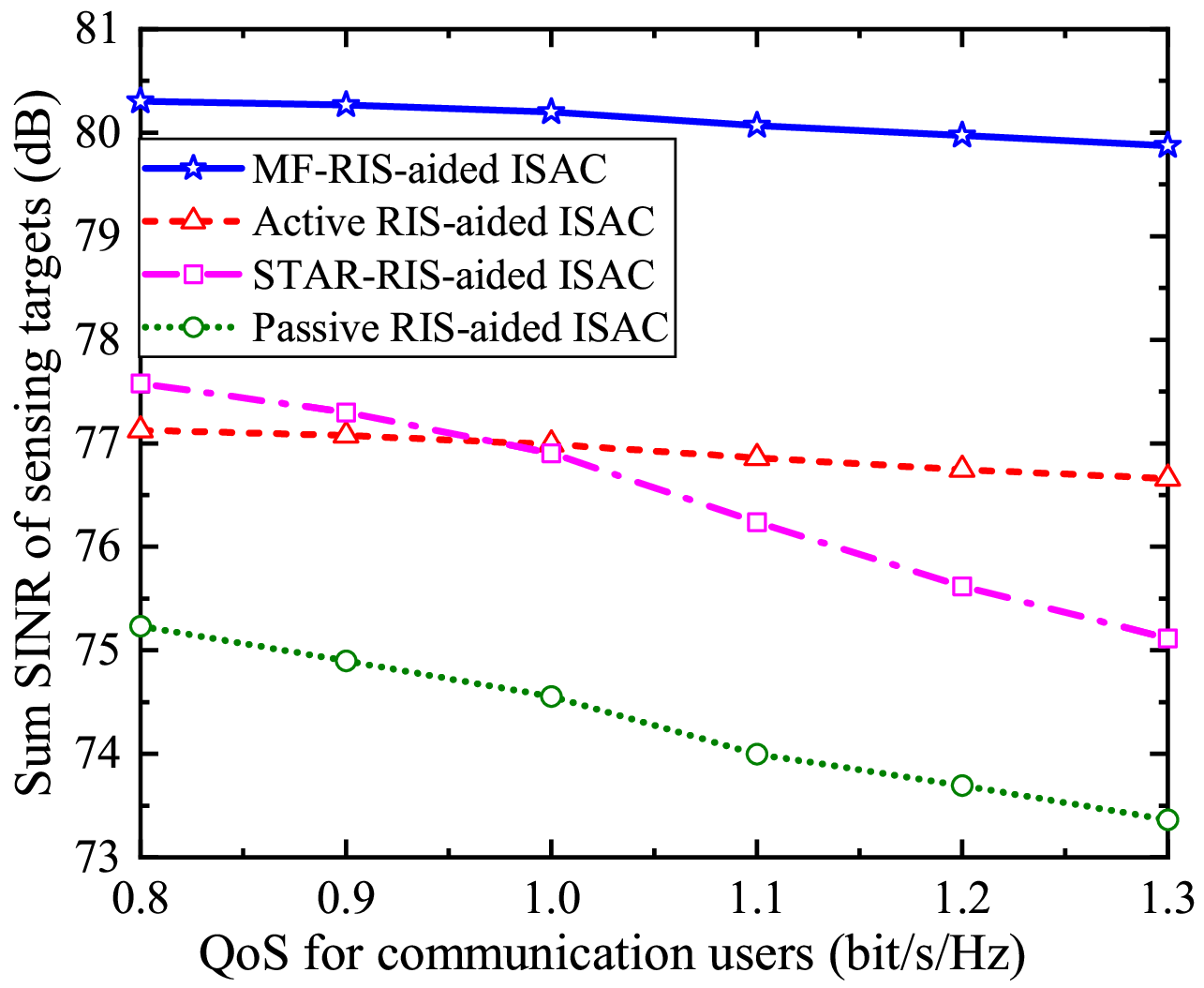}
			\end{minipage}
		}  \hspace{5 mm}
		\subfloat[Sum SINR of sensing targets versus the number of RIS elements]{\label{Fig5_1}
			\begin{minipage}[t]{0.45 \textwidth}
				\centering
				\includegraphics[width= 3.2 in]{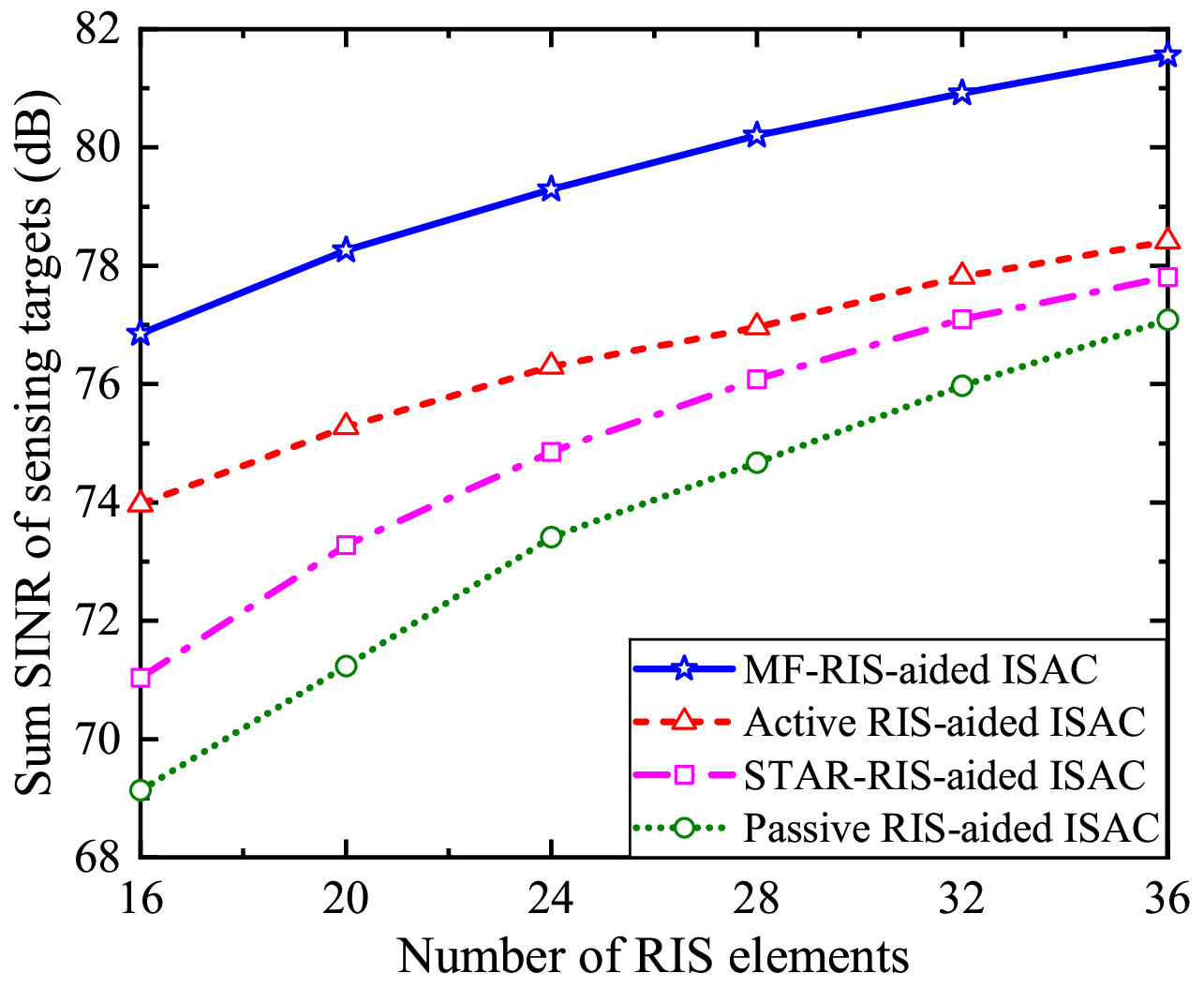}
			\end{minipage}
		}
		\caption{Performance evaluation of ISAC systems aided by different RIS types: (a) sum SINR of sensing targets versus the quality-of-service (QoS) of communication users, and (b) sum SINR of sensing targets versus the number of RIS elements. Here, the QoS is measured by the spectral efficiency.}
		\label{Fig5}
	\end{figure*}
  
	\section{Numerical Results}
	In this section, we present numerical results to demonstrate the performance of our proposed MF-RIS-aided ISAC system, where a DFRC BS is considered to serve four communication users and four sensing targets.
	The distance between the BS to the MF-RIS is 50 meters (m).
	Additionally, all users and targets are randomly dispersed around the BS within a radius of 100 m.
	The numbers of transmit and receive antennas at the BS and users are set as $N_t = N_r = 8$.
	The total power budget of the BS and the MF-RIS is $P_{\max} = 25 {\rm \ dBm}$.
	The signal-to-interference-plus-noise ratio (SINR) and mean-square-error (MSE) thresholds of communication users and sensing targets are set to $10 {\rm \ dB}$ and $0.25$, respectively.
	We adopt passive RIS, STAR-RIS, and active RIS as three benchmarks for performance comparison in the considered ISAC systems.

	Fig. \ref{Fig5_2} shows the sum SINR of sensing targets versus the QoS of communication users.
	It is observed that as the QoS requirement for communication increases, the sum SINR for sensing decreases for all schemes.
	This result is expected as there exists an inherent competition between communication and sensing goals in terms of resource allocation.
	Specifically, increasing the communication rate often requires more spectrum resources and energy consumption, which leads to less resources available for sensing.
	However, compared to the benchmarks, the proposed MF-RIS exhibits best performance.
	This confirms that MF-RISs can achieve better overall system performance by flexibly engineering its reflective, refractive, and amplifying properties to better trade off communication and sensing resource allocation.	
	Another interesting phenomenon is that the sensing performance of active RIS types (active RISs and MF-RISs) is less affected by changes in the required communication performance compared to passive RIS types (passive RISs and STAR-RISs). 
	This is because active RIS types are able to flexibly manage the RIS available power to compensate for communication performance on-demand.

	Fig. \ref{Fig5_1} depicts the sum SINR of the sensing targets versus the number of RIS elements.
	We observe that due to the increased number of elements providing more optimization DoFs, all curves exhibit an increasing trend.
	Moreover, MF-RISs always outperform the benchmarks, while passive RISs appear to be the worst. 
	This is because passive RISs, STAR-RISs, and active RISs can all be regarded as special cases of MF-RISs, and passive RIS are further special cases of STAR-RISs and active RISs.
	Among two dual-functional RISs, the superiority of active RISs over STAR-RISs evidently demonstrates that signal amplification directly and efficiently boosts DFRC signal reception compared to omnidirectional signal emission.
	The reasons why MF-RISs outperform active RISs are twofold:
	\textcircled{\scriptsize{1}} Active RIS achieves full-space coverage by deploying a reflective RIS and a refractive RIS, with coefficient adjustments based on fixed elements. 
	\textcircled{\scriptsize{2}} In contrast, MF-RIS adjusts phase shift and amplitude coefficients on a per-element basis, allowing it to leverage the DoFs of each element to manipulate DFRC signals and enhance signal reception across the entire space.
	However, as the RIS scale increases, the growth rate of the performance gain slows down due to the limitation of the total RIS amplification power.
	The plot also shows that the performance gap between the MF-RIS-aided ISAC system and the other types of RIS-aided systems grows larger as the number of RIS elements increases.
	
	\section{Conclusions and Future Work}
	In this article, we integrated MF-RISs into 6G networks to enhance communication and sensing capabilities.
	First, MF-RISs were used to intelligently manipulate the wireless environment to optimize user experience and data transmission efficiency across unicast, multicast, and broadcast systems. 
	Additionally, we designed MF-RIS-aided multiple access technologies to improve spectral efficiency and support large-scale connectivity.
	Furthermore, we discussed the synergistic relationship between MF-RISs and wireless sensing.
	Moreover, we presented cooperation schemes for MF-RIS-aided RCC and DFRC systems, highlighting the advantages of MF-RISs in improving communication and sensing performance. 
	Subsequently, we advocated for multi-objective optimization approaches in MF-RIS-aided ISAC systems to balance various trade-offs. 
	Finally, we provided numerical results to validate the effectiveness of MF-RISs in enhancing ISAC performance.
	In the following, we outline several unresolved open problems that warrant further exploration in future work.

	\subsubsection{MF-RIS-Aided Near-Field ISAC}
	First of all, unlike plane waves commonly encountered in far-field communications, spherical waves in near-field ISAC inherently have a curved wavefront that affects how electromagnetic waves propagate and interact with objects in the near-field region.
	Thus, MF-RISs need to be designed and configured to accurately manipulate these curved wavefronts, which requires sophisticated algorithms and control strategies.
	Second, the proximity of the target to the sensing device necessitates high spatial resolution and precise beamforming.
	The rapid variation of spherical waves across space make it difficult to focus energy accurately on specific targets or regions.
	Therefore, MF-RISs must be able to dynamically adjust its elements to compensate for these effects and create beams tailored to the specific sensing or communication needs.

	\subsubsection{AI-Based Beamforming Design for MF-RIS-Aided ISAC}
	Usually, ISAC systems operate in dynamic environments with moving targets, users, and electromagnetic interference.
	Moreover, the additional DoF introduced by MF-RISs makes accurate channel estimation a challenging task.
	Integrating artificial intelligence (AI) into MF-RIS-aided ISAC systems may opens up new possibilities for enhancing the performance and adaptability by training AI model on large datasets.
	Furthermore, AI-driven optimization frameworks can balance the competing objectives of sensing accuracy and communication throughput, finding Pareto-optimal solutions that satisfy both requirements.
	Overall, AI has the potential to revolutionize MF-RIS-aided ISAC systems by enabling more intelligent and adaptive channel estimation, beamforming, and resource allocation strategies.

	\subsubsection{MF-RIS Integrated Sensing, Communication, and Computing}
	The design of MF-RIS-aided systems should not be limited to RF communications.
	Besides, integrating MF-RISs with intelligent computing and sensing capabilities can enable more comprehensive decision-making in complex scenarios.
	Achieving optimal multi-objective performance in an integrated system requires a holistic approach.
	Therefore, it is essential to explore intelligent surface co-design approaches that consider the interplay between the communication environment, computation tasks, and sensing applications.
	Meanwhile, investigating cross-domain optimization techniques that can jointly improve communication quality, computing efficiency, and sensing accuracy is also an important research direction.

\bibliographystyle{IEEEtran}
\bibliography{IEEEabrv,ref}

\end{document}